\documentclass[pra,aps,twocolumn,showpacs]{revtex4}
\usepackage{amsmath,graphicx,epsfig}

\def\ket#1{{|#1\rangle}}

\def\threej(#1,#2)(#3,#4)(#5,#6){\begin{pmatrix}#1&#3&#5\\#2&#4&#6\end{pmatrix}}
\def\sixj(#1,#2,#3)(#4,#5,#6){\begin{Bmatrix}#1&#2&#3\\#4&#5&#6\end{Bmatrix}}
\def\ninej(#1,#2,#3)(#4,#5,#6)(#7,#8,#9){\begin{Bmatrix}#1&#2&#3\\#4&#5&#6\\#7&#8&#9\end{Bmatrix}}

\def\cg(#1,#2)(#3,#4)(#5,#6){{\langle#1,#2,#3,#4}\ket{#5,#6}}

\def\fig_width{3. in} 
\draft

\newlength{\defbaselineskip}
\newcommand{\setlinespacing}[1]%
           {\setlength{\baselineskip}{#1 \defbaselineskip}}

\begin{document}

\title{Spin-exchange relaxation free magnetometry with Cs vapor} 

\author{M.\ P.\ Ledbetter}\email{ledbetter@berkeley.edu}
\author{I.\ M.\ Savukov}
\author{V.\ M.\ Acosta}
\author{D.\ Budker}\email{budker@berkeley.edu}
\affiliation{Department of Physics, University of California at
Berkeley, Berkeley, California 94720-7300}
\author{M.\ V.\ Romalis}
\affiliation{Department of Physics, Princeton University, Princeton,
New Jersey, 08544}

\date{\today}


\begin{abstract}
We describe a Cs atomic magnetometer operating in the spin-exchange
relaxation-free (SERF) regime.  With a vapor cell temperature of
$103^\circ\rm{C}$ we achieve intrinsic magnetic resonance widths
$\Delta B=17~{\rm \mu G}$ corresponding to an electron
spin-relaxation rate of $300~{\rm s^{-1}}$ when the spin-exchange
rate is $\Gamma_{SE}=14000~ {\rm s^{-1}}$. We also observe an
interesting narrowing effect due to diffusion. Signal-to-noise
measurements yield a sensitivity of about $400\thinspace{\rm
pG/\sqrt{Hz}}$. Based on photon shot noise, we project a sensitivity
of $40~{\rm pG/\sqrt{Hz}}$. A theoretical optimization of the
magnetometer indicates sensitivities on the order of $2~{\rm
pG/\sqrt{Hz}}$ should be achievable in a $1~{\rm cm^3}$ volume.
Because Cs has a higher saturated vapor pressure than other alkali
metals, SERF magnetometers using Cs atoms are particularly
attractive in applications requiring lower temperatures.

\end{abstract}




\maketitle

\section{Introduction}
Sensitive atomic magnetometers have recently found application in
the field of magnetic resonance imaging \cite{Xu2006},
magneto-encephalography \cite{Xia2006}, and searches for physics
beyond the standard model \cite{Bear2000}. A recent review may be
found in Ref. \cite{BudkerRomalis2007}. The most sensitive atomic
magnetometers presently are the spin-exchange relaxation free (SERF)
magnetometers \cite{Kominis03} in which relaxation due to
spin-exchange collisions is eliminated by operating in the regime
where the spin-exchange rate is much greater than the rate of Larmor
precession \cite{Happer73,Happer77}.  In Ref. \cite{Kominis03},
sensitivity of $5~{\rm pG/\sqrt{Hz}}$ was achieved with the cell
operating at $190^\circ\thinspace{\rm C}$ using potassium atoms.
Estimates of the fundamental sensitivity limit of such magnetometers
are several orders of magnitude better for a ${\rm 1~ cm^3}$ volume
and scale as the square-root of the spin-destruction cross section.

Here we demonstrate operation of a Cs magnetometer in the SERF
regime, achieving a sensitivity of about $400\thinspace{\rm
pG/\sqrt{Hz}}$ with a vapor cell temperature of only
$103^\circ\thinspace{\rm C}$.  The overlapping volume of the pump
and probe beams is about $0.02~{\rm cm^3}$, but the effective
volume, determined by diffusion, is about $1~{\rm cm^3}$. Based on
optical rotation measurements, the projected photon shot noise limit
for our experimental conditions is about $40~{\rm pG/\sqrt{Hz}}$.
The spin-destruction cross section for Cs is $\sigma_{SD}=2\times
10^{-16}\thinspace{\rm cm^2}$, about 200 times larger than for K
\cite{Bhaskar1980}. Hence, the fundamental sensitivity of a  Cs SERF
magnetometer should be roughly a factor of 14 worse than one based
on K. However, it is often the case that environmental noise due to,
for example, Johnson currents in magnetic shields is far larger than
atomic shot noise, and hence little is lost by using Cs. One of the
primary motivations for investigating Cs in the SERF regime is the
fact that it has the highest saturated vapor pressure of all the
stable alkalis, yielding significantly lower operating temperatures.
This opens up the possibility of operating in the SERF regime with
paraffin coated cells, of interest due to the lower light power
requirements in evacuated cells. The low temperature aspect of Cs is
also attractive for applications such as NMR measurements with
liquids in microfluidic channels, which is expected to be an
important measurement modality in future ``lab-on-a-chip'' devices
\cite{Microfluidicref}.

\section{Bloch equations}

A full treatment of the system requires the use of density matrix
theory (see, for example, Ref \cite{Appelt98}). However, the
description can be greatly simplified when the spin-exchange rate
\begin{equation}\label{Eq:Spinexrate}
\Gamma_{SE}=T_{SE}^{-1}=n\sigma_{SE} \bar{v}
\end{equation}
(here $n$ is the alkali number density, $\sigma_{SE}\approx 2\times
10^{-14}\thinspace{\rm cm^{-2}}$ is the spin-exchange cross section,
and $\bar{v}$ is the average relative velocity of the colliding
alkali atoms) is much faster than precession in the magnetic field,
$\Gamma_{SE}\gg g_s\mu_B B/(2I+1)$. Here $g_s\approx 2$ is the
electron Land\'e factor, $\mu_B$ is the Bohr magneton, $B$ is the
magnitude of an applied magnetic field, and $I$ is the nuclear spin.
In this case the density matrix assumes a spin-temperature
distribution and the ground state can be well described by Bloch
equations for the electron spin polarization $\mathbf{P}=\langle
\mathbf{S} \rangle/S$ \cite{Allred02,Seltzer04,Kornack02}:
\begin{eqnarray}
\nonumber\lefteqn{\frac{d \mathbf{P}}{dt}=\frac{1}{q(P)}\times}\\
    & &\left( g_s\mu_B \mathbf{P}
    \times
    \mathbf{B}+R(\mathbf{s}-\mathbf{P})-\frac{\mathbf{P}}{T_1,T_2}-\Gamma_{pr}
    \mathbf{P}\right).\label{Eq:BlochEqns}
\end{eqnarray}
Here $\mathbf{s}$ is the optical pumping vector along the direction
of propagation of the pump with magnitude equal to the degree of
circular polarization, $R$ is the optical pumping rate due to the
pump beam, and $\Gamma_{pr}$ is the rate of depolarization due to
the linearly polarized probe beam. The quantity $q(P)$ is the
nuclear slowing-down factor, which for nuclear spin $I=7/2$, is
\cite{Savukov2005PRA}
\begin{equation}\label{Eq:Slowingdown}
    q(P)=\frac{2 \left(P^6+17 P^4+35 P^2+11\right)}{P^6+7 P^4+7
    P^2+1}.
\end{equation}
In the low polarization limit $q(0)=22$, while in the high
polarization limit, $q(1) = 8$. The latter limit, when all atoms are
pumped into the stretched state, corresponds to the slowing down
factor for nuclear spin $I=7/2$ in the absence of spin-exchange
collisions $q=2I+1=8$. In Eq. \eqref{Eq:BlochEqns}, $T_1$ and $T_2$
are the longitudinal and transverse (with respect to $\mathbf{B}$)
relaxation times, respectively. The transverse relaxation time can
be written
\begin{equation}\label{Eq:T_2}
    \frac{1}{T_2}=\Gamma_{SD}+\frac{1}{T_2^{SE}},
\end{equation}
where $\Gamma_{SD}$ is the electron spin-destruction rate and
$(T_2^{SE})^{-1}$ is the contribution to relaxation from
spin-exchange collisions. For low polarizations and small magnetic
fields, relaxation due to spin-exchange is quadratic in the magnetic
field \cite{Happer77}
\begin{equation}\label{Eq:spinexrelax}
    \frac{1}{T_2^{SE}}=\frac{\Omega_0^2}{\Gamma_{SE}}\frac{q(0)^2-(2I+1)^2}{2},
\end{equation}
where $\Omega_0 = B g_s \mu_B/q(0)$.

In some of the measurements described below, small, quasistatic
magnetic fields are applied, and the conditions are such that
relaxation due to spin-exchange collisions can be ignored.  In our
present experimental setup (see Fig. \ref{Fig:exp_setup}) optical
pumping is along the $z$ axis (the longitudinal direction) and
optical rotation of the probe is due to $P_x$. The steady state
solutions to Eq. \eqref{Eq:BlochEqns} can be found by setting the
L.H.S. to zero, resulting in
\begin{eqnarray}
   P_x  &=& P_0\frac{B_x B_z-B_y\Delta B}{ B^2+\Delta B^2}, \label{Eq:Sxsoln}\\
   P_z &=& P_0 \frac{B_z^2+\Delta B^2}{B^2+\Delta_B^2},
   \label{Eq:Szsoln}
\end{eqnarray}
where
\begin{eqnarray}
  \Delta B &=& (R+\Gamma_{pr}+\Gamma_{SD})/g_s \mu_B\,,  \label{Eq:DeltaB}\\
  P_0 &=& sR/(R+\Gamma_{pr}+\Gamma_{SD}) \label{Eq:S0}.
\end{eqnarray}
Note that Eqs. \eqref{Eq:Sxsoln}-\eqref{Eq:S0} are independent of
the nuclear spin.

To study the effects of spin exchange, we find it convenient to
apply a small rotating field $\mathbf{\hat{x}}B_1 \sin{\omega
t}+\mathbf{\hat{y}} B_1 \cos{\omega t}$ in the presence of a larger
bias field $B_z$. In this case we find the in-phase and quadrature
components of $P_x$
\begin{eqnarray}
   P_x^{(in)}  &=& -P_0\frac{g_s\mu_B B_1}{q(P)}
   \frac{\Delta\omega}{(\omega-\Omega_0)^2+\Delta\omega^2}\,,\label{Eq:rotfieldsoln1} \\
   P_x^{(out)}  &=& -P_0\frac{g_s\mu_B B_1}{q(P)}
  \frac{\omega-\Omega_0}{(\omega-\Omega_0)^2+\Delta\omega^2}\,,\label{Eq:rotfieldsoln2}
\end{eqnarray}
where $\Delta\omega=(R+\Gamma_{pr}+\Gamma_{SD}+1/T_2^{SE})/q(P)$.

In the presence of rapid quenching of the excited state by N$_2$,
the effects of optical pumping and the optical properties of the
medium can be treated with an effective ground state formalism
\cite{Happer1967}.  When the pressure broadened optical width is
much larger than the hyperfine splitting, the optical pumping rate
for light of frequency $\nu$ is given by \cite{Appelt99}
\begin{equation}\label{Eq:oprate}
    R = \Phi \sigma = \Phi r_e c
    f\frac{\Delta \nu/2}{(\Delta\nu/2)^2+(\nu-\nu_0)^2},
\end{equation}
where $\sigma$ is the absorption cross section, $r_e=2.8\times
10^{-13}~{\rm cm}$ is the classical radius of the electron, $c$ is
the speed of light, $f$ is the oscillator strength (roughly 1/3 for
D1 light and 2/3 for D2 light), $\Phi$ is the photon flux per unit
area and $\Delta\nu$ is the full-width at half-maximum of the
optical transition of frequency $\nu_0$. Optical rotation of
linearly polarized D2 light, propagating in the $x$ direction is
dispersive in the detuning of the probe beam from optical resonance
\cite{Happer1967,Kornackthesis},
\begin{equation}\label{Eq:optrotation}
    \phi = \frac{1}{4}lr_e c f n P_x D(\nu),
\end{equation}
where $l$ is the optical path length and
$D(\nu)=(\nu-\nu_0)/[(\nu-\nu_0)^2+(\Delta\nu/2)^2]$.

\section{Experimental setup and procedure}

\begin{figure}
  \includegraphics[width=3.4 in]{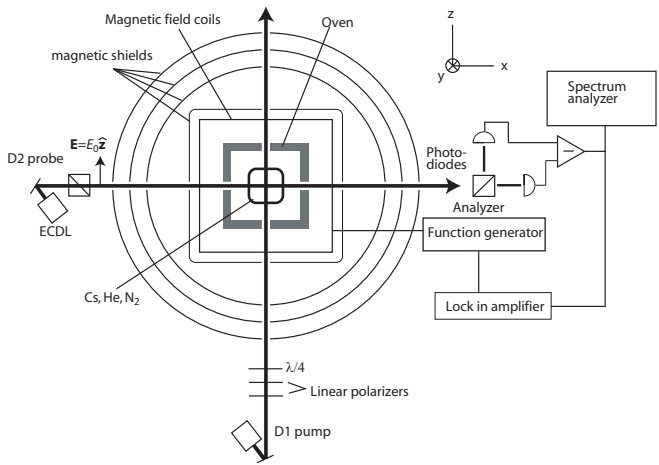}\\
  \caption{Experimental setup.  A four-layer magnetic shield provides a magnetically shielded
  environment. Circularly polarized light tuned to the D1 line, propagating in the $z$ direction
  produces ground state orientation in the $z$ direction.  The $x$ component of orientation $S_x$
  is detected via optical rotation of linearly polarized light, tuned to the D2 line, propagating in the
  $x$ direction.}\label{Fig:exp_setup}
\end{figure}

The experimental setup is shown in Fig. \ref{Fig:exp_setup}. A glass
cell containing a droplet of Cs metal, 600 Torr He buffer gas (to
reduce the rate at which atoms in the central part of the cell
diffuse to the cell walls) and 20 Torr of ${\rm N_2}$ (to eliminate
radiation trapping and improve optical pumping efficiency) is placed
inside a four-layer set of magnetic shields. The cell has a roughly
cubic profile, about 2 cm on a side. After degaussing, the residual
fields inside the magnetic shields are on the order of $2-3~{\rm \mu
G}$. From the known rates of pressure broadening of Cs lines by
helium \cite{Andalkar2002b}, we extrapolate the FWHM of the D1 and
D2 optical resonances to be $\Delta \nu = 15.7\thinspace{\rm and}
\thinspace 14.1~{\rm GHz}$, respectively. The cell was heated to
${\rm 103^\circ C}$ by flowing hot air through the space between the
walls of a double-wall oven. The oven was designed so that the
optical path was unperturbed by the flowing air. At $103^\circ{\rm
C}$, the saturated Cs vapor concentration is about $[{\rm
Cs}]=1.7\times 10^{13}~{\rm cm^{-3}}$. Optical pumping was
accomplished by circularly polarized laser light propagating in the
$z$ direction tuned to the center of the Cs D1 line (the exact
tuning was chosen to minimize light shifts). The pump beam was about
4 mm in diameter. A linearly polarized probe beam, with cross
section $\approx 2 \times 3\thinspace{\rm mm^2}$, tuned about 5
optical linewidths from the center of the pressure broadened D2 line
(where signal was maximized) propagated in the $x$ direction through
the cell.  Note that based on Eq. \eqref{Eq:optrotation}, one
\begin{figure}
  \includegraphics[width=3.4 in]{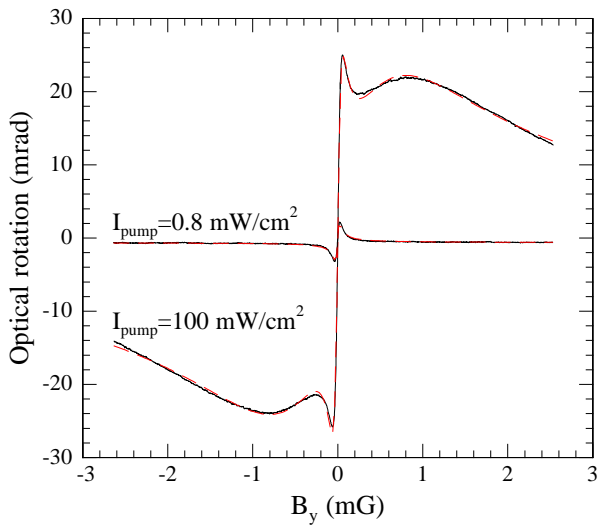}\\
  \caption{Optical rotation (solid line) of a weak ($\approx 0.5~{\rm mW/cm^2}$) probe beam
  as a function of $B_y$ for pump intensities as indicated next to each trace. The dashed lines
  are fits to one or two dispersive Lorentzians, as described in the text.}\label{Fig:Byscan}
\end{figure}
expects the maximum optical rotation to occur for detuning from
resonance by $\Delta \nu/2$.  However, the signal is the product of
the transmission and optical rotation, and since there are many
optical depths in our cell (on resonance, 1 cm corresponds to about
12 optical depths under the conditions of our measurements), it was
necessary to detune far from resonance. Circular birefringence of
the medium proportional to $P_x$ rotates the polarization of the
probe beam, which is analyzed after the cell with a balanced
polarimeter. To investigate the zero-field resonance, where behavior
is described by Eq. \eqref{Eq:Sxsoln}, optical rotation of the probe
beam was measured as a function of a static field $B_y$ for all
other fields zeroed. $B_x$ ($B_z$) can be zeroed by making use of
Eq. \eqref{Eq:Sxsoln}: a small, slowly oscillating field ($B_x$)
$B_z$ is applied while $B_x$ ($B_z$) is adjusted until the resulting
signal is zero.  To study the effects of spin-exchange broadening at
non-zero fields, optical rotation was detected synchronously using a
lock-in amplifier while a small rotating magnetic field
$\mathbf{\hat{y}}B_1\cos{\omega t}+\mathbf{\hat{x}}B_1\sin{\omega
t}$ was applied in the presence of a larger bias field $B_z$.

\section{Experimental results and discussion}

\subsection{Zero-field resonance}

\begin{figure}
  \includegraphics[width=3.4 in]{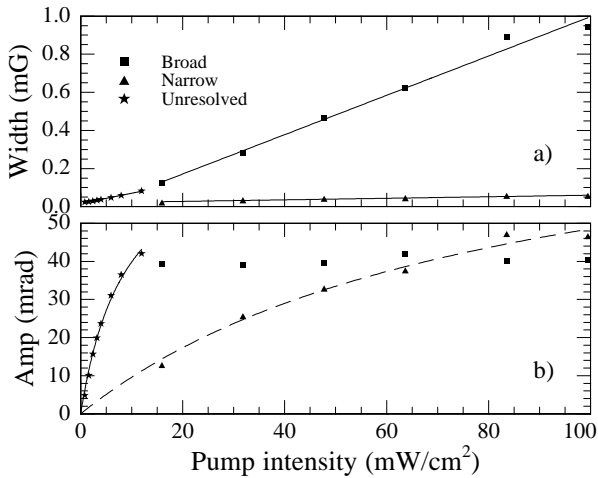}\\
  \caption{Zero-field resonance half-width at half-max (top panel) and peak-to-peak amplitude (bottom panel) as a function of
  pump power. The lines overlaying the data are fits described in the text.}\label{Fig:Byscan_params}
\end{figure}

In Fig. \ref{Fig:Byscan} the solid curves show the magneto-optical
rotation of a weak probe beam ($I_{pr}\approx 0.5~{\rm mW/cm^2}$) as
a function of magnetic field $B_y$ for two different pump
intensities. The dashed line overlaying the data for weak pump
light, $I_{pump}=0.8~{\rm mW/cm^2}$, is a fit based on Eq.
\eqref{Eq:Sxsoln} with $\Delta B = 23~{\rm \mu G}$. For more intense
pump light $I_{pump}=100~{\rm mW/cm^2}$, the optical rotation is
well described by the sum of two dispersive Lorentzians with widths
$\Delta B = 56~{\rm \mu G}$ and $\Delta B = 940~{\rm \mu G}$, as
indicated by the dashed red line. In Fig. \ref{Fig:Byscan_params} we
plot the width and peak-to-peak amplitude of the optical rotation
for the single feature observed for pump intensities below about
$15~{\rm mW/cm^2}$ (stars) and for the nested features observed at
higher pump intensities (squares and triangles).  In either regime,
the width is linear in light intensity.  When the two features
become resolved, the amplitude of the broad resonance appears
saturated, while the amplitude of the narrow resonance continues to
grow, approaching saturation at the highest light power.

The difference in the behavior of magneto-optical rotation for low
and high pump intensities is essentially due to diffusion of atoms
into and out of the pump beam, creating two regions with differing
polarization and rates of power broadening. Similar narrowing
effects due to diffusion have been observed in the context of
electromagnetically induced transparency in buffer gas cells
\cite{Xiao2006}. Nested resonances have also been observed in the
context of nonlinear magneto-optical rotation in paraffin coated
cells \cite{Budker98}. In a simplified model, the cell can be
divided into regions illuminated and dark with respect to the pump
beam. The optical pumping rate $R$ inside the pump beam is
determined by Eq. \eqref{Eq:oprate} so that the power broadened
width and polarization of atoms inside the pump beam is given by
Eqs. \eqref{Eq:DeltaB} and \eqref{Eq:S0}. The region outside the
pump beam is polarized via diffusion and hence the average optical
pumping rate is suppressed by the ratio
$\Gamma_{SD}/(\Gamma_{SD}+\Gamma_D)$, where $\Gamma_D$ is the rate
of diffusion of atoms into and out of the pump beam. Hence the width
of the zero field resonance and polarization for atoms outside the
pump beam is
\begin{eqnarray}
    \Delta B^{out} &=&\label{Eq:deltaBout}
    \frac{1}{g_s\mu_B}\left(R\frac{\Gamma_{SD}}{\Gamma_{SD}+\Gamma_D}+\Gamma_{SD}\right),\\
    P_0^{out} &=& \frac{R}{R+\Gamma_D+\Gamma_{SD}}.\label{Eq:P0out}
\end{eqnarray}
For pump intensities such that $R$ is small compared to
$\Gamma_{SD}$ the width of the resonance due to atoms inside and
outside of the pump beam is determined primarily by $\Gamma_{SD}$
and hence a single feature is visible. The polarization of the
region outside the pump beam is smaller than inside the pump beam by
a factor $\Gamma_{SD}/(\Gamma_{SD}+\Gamma_D)$ and hence the
amplitude of optical rotation is dominated by the region inside the
pump beam. When the optical pumping rate inside the beam is large
compared to $\Gamma_{SD}$, the region inside the pump beam becomes
saturated and optical rotation is sufficiently broadened so that the
smaller narrow feature due to the region outside the pump beam
becomes distinguishable. As $R$ becomes large compared to
$\Gamma_D+\Gamma_{SD}$, the region outside the pump becomes fully
polarized and optical rotation approaches an asymptote.

Overlaying the data in Fig. \ref{Fig:Byscan_params}a are linear fits
based on Eq. \eqref{Eq:DeltaB} with $R=\eta I_{pump}$.  The fit to
the low pump intensity data yields $\eta=93~{\rm s^{-1}/(mW/cm^2)}$
and zero light-power width $\Delta B_0=17\pm 3\thinspace{\rm \mu G}$
corresponding to $\Gamma_{SD}=300\pm52~{\rm s^{-1}}$. This can be
compared to the expected spin-destruction rate based on previous
measurements of the spin-destruction cross sections
\begin{equation}\label{Eq:relaxation}
\Gamma_{SD}= \lbrack{\rm Cs}\rbrack\bar{v}^{\rm Cs}\sigma_{SD}^{\rm
Cs}+\lbrack{\rm He}\rbrack\bar{v}^{\rm He}\sigma_{SD}^{\rm
He}+\lbrack{\rm N_2}\rbrack\bar{v}^{\rm N_2}\sigma_{SD}^{\rm
    N_2}.
\end{equation}
Here, $\sigma_{SD}^{\rm Cs}=2\times 10^{-16}\thinspace{\rm cm^2}$
\cite{Bhaskar1980}, $\sigma_{SD}^{\rm He}=3\times
10^{-23}\thinspace{\rm cm^2}$ \cite{Beverini1971}, and
$\sigma_{SD}^{\rm N_2}=6\times 10^{-22}\thinspace{\rm cm^2}$
\cite{Beverini1971} are the spin destruction cross sections for
Cs-Cs, Cs-He and Cs-$\rm{N_2}$ collisions respectively. The mean
relative velocity $\bar{v}^{\rm X}$ differs between colliding pairs,
and hence the superscript in Eq. \eqref{Eq:relaxation}. The
contributions to the spin-destruction rate from Cs, He and ${\rm
N_2}$ collisions are $119\thinspace{\rm s^{-1}}$, $88\thinspace{\rm
s^{-1}}$ and $32\thinspace{\rm s^{-1}}$ respectively yielding a
total spin destruction rate $\Gamma_{SD}=240~{\rm s^{-1}}$, in
reasonable agreement with the present measurements.

In Fig. \ref{Fig:Byscan_params}b, the solid line overlaying the
amplitude of the zero-field resonance for low pump intensity is a
fit to $a\eta I/(\eta I+\Gamma_{SD})$ with $\Gamma_{SD}$ fixed by
the value determined above, yielding $a=75~{\rm mrad}$ and $\eta =
33~{\rm s^{-1}/(mW/cm^2)}$, about a factor of 3 smaller than the
pump rate determined from the broadening. The reason for the
difference in pump rates determined from the amplitude and the rate
of broadening is due to diffusion as discussed above: The pump rate
determined by the broadening will be dominated by the illuminated
region, while the pump rate determined from the amplitude represents
an average pump rate over the volume the polarized atoms occupy
during the course of one relaxation period.

\begin{figure}
  \includegraphics[width=3.4 in]{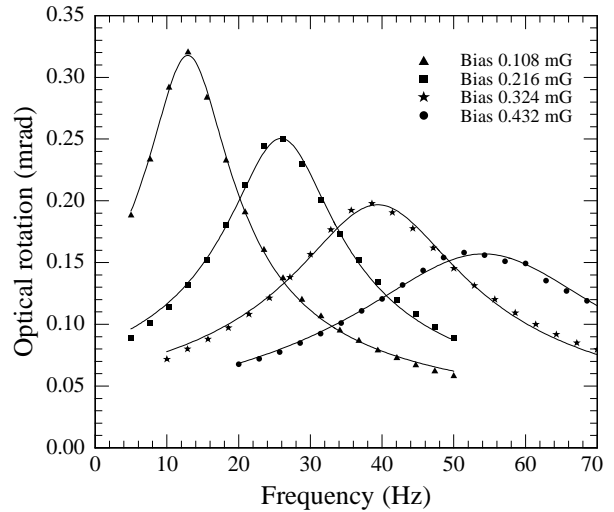}\\
  \caption{Response of magnetometer to a small rotating magnetic field of magnitude
  $B_1=0.55\thinspace{\rm \mu G}$ transverse
  to a larger bias field $B_z$ for several different values of $B_z$.  For these data
  the pump and probe intensities were ${\rm~4 mW/cm^2}$ and ${\rm~1.3 mW/cm^2}$ respectively.}\label{Fig:biasfieldresponse}
\end{figure}

\subsection{Spin-exchange effects}

To explore the effects of spin-exchange on transverse relaxation we
apply a bias magnetic field in the $z$ direction (along the pump
beam) and a small transverse rotating magnetic field to excite a
component of polarization transverse to the bias field. In-phase and
quadrature components of the resulting optical rotation signal are
detected synchronously using a lock-in amplifier.  In Fig.
\ref{Fig:biasfieldresponse} we show the quadrature sum of the in-
and out- of phase optical rotation signals as a function of
frequency for several different values of the bias magnetic field.
Overlaying the data are fits to
$a\Delta\omega/\sqrt{(\omega-\Omega_0)^2+\Delta\omega^2}$ (see Eqs.
\eqref{Eq:rotfieldsoln1} and \eqref{Eq:rotfieldsoln2}). For these
data, the pump and probe intensity were about 4 and 1.3 ${\rm
mW/cm^2}$, respectively. Based on the data shown in Fig.
\ref{Fig:Byscan_params}, these intensities produce power broadening
by about a factor of 2 over the zero light-power width, however the
slowing down factor, determined from a linear fit to $\Omega_0$, was
very nearly $q=22$ indicating that the polarization was quite low
(see Eq. \eqref{Eq:Slowingdown}).

\begin{figure}
  \includegraphics[width=3.4 in]{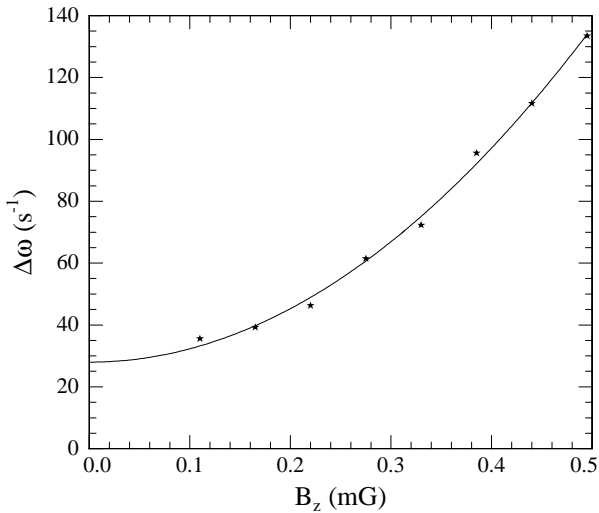}\\
  \caption{Half-width at half-maximum of the bias-field resonances shown in Fig. \ref{Fig:biasfieldresponse}
  as a function of magnetic field.}\label{Fig:width_gyromagneticratio}
\end{figure}

In Fig. \ref{Fig:width_gyromagneticratio} we plot the half-width at
half-maximum $\Delta\omega$ of the resonances shown in Fig.
\ref{Fig:biasfieldresponse} as a function of the bias field.
Overlaying the data is a fit based on Eq. \eqref{Eq:spinexrelax}
with $\Delta\omega = (q(0) T_2)^{-1}$ and $\Omega_0 = g_s \mu_B
B/q(0)$, allowing for a constant offset due to spin-destruction
collisions, diffusion and power broadening,
yielding a spin-exchange rate $\Gamma_{SE}=14300\pm 350~{\rm
s^{-1}}$. For $n=1.7\times 10^{13}~{\rm cm^{-3}}$ obtained from the
saturated vapor pressure curve, Eq. \eqref{Eq:Spinexrate} gives a
spin-exchange rate of about $12000~{\rm s^{-1}}$.  Temperature
fluctuations of 2 or 3 degrees could cause significant variations in
the vapor pressure, and hence we consider this reasonable agreement
with measurements of spin-exchange broadening.

\subsection{Sensitivity}
\begin{figure}
  \includegraphics[width=3.4 in]{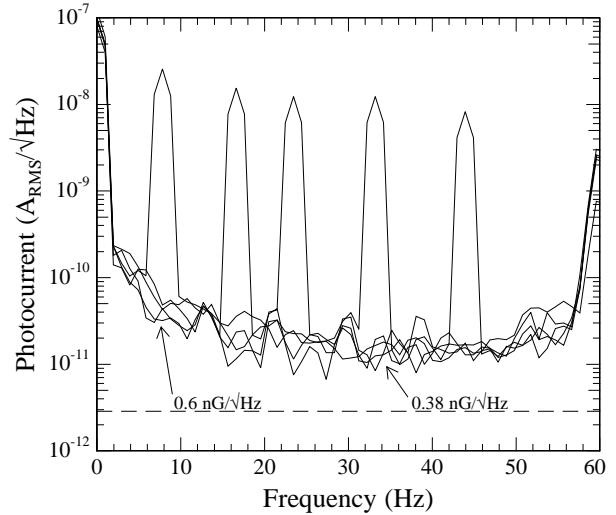}\\
  \caption{Fourier transform of magnetometer signal (solid lines) with calibration peaks of amplitude
  $B_1 = 0.55~{\rm \mu G}$ applied at several different frequencies.
 The dashed line represents photon shot noise.}\label{Fig:Sensitivity}
\end{figure}

We evaluate the performance of the magnetometer by monitoring the
noise level at the output of the balanced polarimeter using a
Stanford Research Systems SR770 spectrum analyzer.  To calibrate the
magnetometer, we apply a small oscillating field $B_y = B_1
\cos\omega t$ with $B_1=0.55~{\rm \mu G}$ at several different
frequencies. The resulting spectra are shown as solid lines in Fig.
\ref{Fig:Sensitivity}.  The triangular shape of the calibration
peaks is due to the use of the built in Hann windowing function with
coarse spectral resolution. For these data, the pump and probe
intensities were $100~{\rm mW/cm^2}$ and $4~{\rm mW/cm^2}$,
respectively, and all three components of the DC magnetic field have
been zeroed. The sensitivity, in ${\rm G_{RMS}/\sqrt{Hz}}$ is
determined by $\delta B = B_1/(\sqrt{2} {\rm S/N})$ yielding a
sensitivity of about $400~{\rm pG_{RMS}/\sqrt{Hz}}$ at 30 Hz and
about $600~{\rm pG_{RMS}/\sqrt{Hz}}$ at 10 Hz.  When the pump beam
was blocked, effectively turning the magnetometer off, noise at low
frequencies dropped by about a factor of 5.  The excess noise could
be due to real fluctuations of the ambient field (either due to
imperfect magnetic shielding or noise in the current source), or due
to fluctuations in the pump power coupled with misalignment of the
pump and probe beams. The dashed line in Fig. \ref{Fig:Sensitivity}
represents the estimated photon shot noise limit in the difference
of the photocurrents for unit bandwidth, $\delta I=\sqrt{4eI}$ where
$I$ is the photocurrent in one channel of the balanced polarimeter,
yielding a photon shot noise limited sensitivity of about $40~{\rm
pG/\sqrt{Hz}}$ at 10 Hz. Optimization of geometry to maximize the
overlapping volumes of the pump and probe beams will likely yield
improvements in the photon shot noise limit.

\section{Conclusions}
We demonstrated a Cs atomic magnetometer in the spin-exchange
relaxation free regime. The primary advantage of using Cs is the
ability to work at lower temperatures.  Future work with atomic
magnetometers in the context of microfluidic NMR will make use of
this feature. At $103^\circ{\rm C}$ we realized magneto-optical
rotation features with intrinsic linewidths of $17~{\rm \mu G}$
corresponding to a relaxation rate of about $300~{\rm s^{-1}}$ when
the spin-exchange rate was about $\Gamma_{SE}=14000~{\rm s^{-1}}$.
We achieved a sensitivity of $400\thinspace{\rm pG/\sqrt{Hz}}$.
Based on estimates of the photon shot noise, we project a
sensitivity of about $40\thinspace{\rm pG/\sqrt{Hz}}$. We suspect
that the demonstrated sensitivity was limited by pump laser noise
and ambient magnetic field noise. Theoretical optimization of the
magnetometer (presented in the Appendix below) indicates it should
be possible to achieve sensitivity on the order of $2~{\rm
pG/\sqrt{Hz}}$ in a $1~{\rm cm^{-3}}$ volume.  We believe the
primary reason for falling short of this level is suboptimal
geometry (probe beam cross section was only $2\times 3~{\rm mm^2}$)
detuning and light power of both pump and probe. This work was
supported by an ONR MURI grant.

\section{Appendix: Theoretical optimization}
We now present a theoretical optimization of the magnetometer,
maximizing sensitivity to small, quasistatic fields. The analysis is
similar to Refs. \cite{Savukov2005b} and \cite{Auzinsh2004}, in that
spin-projection noise and photon shot noise are considered
independently (noise due to light shifts is not considered, because
in principle, it can be eliminated by orthogonality of pump and
probe beams \cite{Auzinsh2004}). Spin-projection noise is typically
written as
\begin{equation}\label{Eq:deltaBnaive}
    \delta B \approx \frac{1}{\gamma\sqrt{NtT_2}},
\end{equation}
where $\gamma$ is the gyromagnetic ratio, $N$ is the number of
atoms, $t$ is the measurement time and $T_2$ is the transverse
relaxation time.  In the SERF regime $T_2=q(P)/\Gamma_{SD}$ and
$\gamma = g_s\mu_B/q(P)$ both depend on the nuclear slowing down
factor. Inserting these expressions into Eq. \eqref{Eq:deltaBnaive},
one might conclude that the atomic shot noise limit scales as
$\sqrt{q(P)}$. However, it turns out that the nuclear slowing down
factor drops out of the problem. The reasons are somewhat subtle, so
we go into some detail.

Spin-projection noise arises (in the present geometry) due to
uncertainty in the $x$ component of angular momentum $F_x$, defined
as $\Delta F_x=\sqrt{\langle F_x^2\rangle-\langle F_x \rangle^2}$.
For a spin temperature distribution with polarization in the $z$
direction, $\rho\propto e^{\beta F_z}$, where $\beta =
\ln(1+P)/(1-P)$ is the spin temperature parameter \cite{Appelt98},
$\langle F_x\rangle=0$. Thus $\Delta F_x=\sqrt{{\rm Tr}( \rho
F_x^2)}$.  Evaluation of this trace results in $\Delta
F_x(P)=\sqrt{q(P)/4}$ per atom with $\rho$ normalized so that ${\rm
Tr} \rho=1$. Assuming that $N=nV$ uncorrelated atoms are involved in
the measurement, the ensemble averaged uncertainty scales as
$1/\sqrt{nV}$,
\begin{equation}\label{Eq:deltaFx}
    \delta F_x(P)=\sqrt{\frac{q(P)}{4 nV}}.
\end{equation}
In the large polarization limit,  $\delta F_x(1)=\sqrt{2/nV}$. This
limit can be obtained from the angular momentum commutation
relations $\lbrack F_x,F_y\rbrack=iF_z$ which yield the minimum
uncertainty $\sqrt{\mid \langle F_z\rangle \mid/2}$.  If all the
atoms are in the stretched state, corresponding to $P=1$, $\langle
F_z\rangle=4$, we have, again assuming uncorrelated atoms, $\delta
F_x=\sqrt{2/nV}$. The uncertainty in the low polarization limit is
somewhat larger, $\delta F_x(0)=\sqrt{11/2nV}$. This limit can also
be verified by noting that for an unpolarized sample $\rho=
\mathbf{1}/(2S+1)(2I+1)$ and ${\rm Tr}(\rho F_x^2)={\rm Tr}(\rho
F_z^2)$. As an aside, we note that the reduction in uncertainty
$\delta F_x$ with increasing polarization only occurs for angular
momentum greater than 1/2.

After measuring continuously for time $t$ long compared to the
lifetime of the polarization, $q(P)/(R+\Gamma_{pr}+\Gamma_{SD})$,
the uncertainty is \cite{Savukov2005b,Gardner}
\begin{eqnarray}\label{Eq:deriv1}
    \langle \delta F_x \rangle_t &=& \delta F_x\sqrt{\frac{2
    q(P)}{(R+\Gamma_{pr}+\Gamma_{SD})t}}\\
    &=& \frac{q(P)}{\sqrt{2 t (R+\Gamma_{pr}+\Gamma_{SD})nV}}.
\end{eqnarray}
In a spin-temperature distribution, the ratio of the total angular
momentum to that stored in the electron is given by $q(P)$, $\langle
\mathbf{F}\rangle =(q(P)/2)\mathbf{P}$ \cite{Appelt98} and thus
\begin{equation}\label{Eq:deltaPx}
     \delta P_x  =\frac{2}{q(P)} \langle \delta
    F_x\rangle_t=\sqrt{\frac{2}{t(R+\Gamma_{pr}+\Gamma_{SD})nV}}.
\end{equation}
From Eq. \eqref{Eq:Sxsoln}, we find for all fields zeroed, that the
uncertainty $\delta B_y$ in a measurement of $B_y$ is related to
flucutations of $P_x$ by
\begin{equation}\label{Eq:deltaBy}
    \delta B_y =
    \frac{R+\Gamma_{pr}+\Gamma_{SD}}{g_s\mu_B}\frac{\delta
    P_x}{P_z}.
\end{equation}
Inserting Eq. \eqref{Eq:deltaPx} into Eq. \eqref{Eq:deltaBy} we find
that the spin-projection noise is
\begin{equation}\label{Eq:atomicshotnoise}
    \delta
    B_{spn}=\frac{1}{g_s\mu_BP_z}\sqrt{\frac{2(R+\Gamma_{pr}+\Gamma_{SD})}{nVt}},
\end{equation}
It is interesting to note that this result is independent of any
nuclear slowing-down factors. Neglecting broadening due to the probe
beam, the minimum value of spin-projection noise
\begin{equation}\label{Eq:dBspnmin}
    \delta B_{spn}^{min}=\frac{3}{g_s\mu_B}\sqrt{\frac{3\Gamma_{SD}}{2nVt}}
\end{equation}
is obtained when $R=2\Gamma_{SD}$.

We now address photon shot noise. To simplify the analysis, we
assume that the volume $V$ occupied by the sample is a cube with
sides of length $l$, fully illuminated by both pump and probe. If
the probe beam is detuned far from resonance so that the medium is
optically thin, optical rotation drops slowly, scaling as
$D(\nu)\approx 1/(\nu-\nu_0)$, compared to absorption which scales
as $1/(\nu-\nu_0)^2$, a very favorable situation. In this case,
photon shot noise in the optical rotation angle is given by $\delta
\phi = 1/2\sqrt{\Phi_0 l^2 t}$ where $\Phi_0$ is the probe photon
intensity and $l^2$ is the cross section of the probe beam.
Combining this with Eqs. \eqref{Eq:optrotation} and
\eqref{Eq:deltaBy}, the photon shot noise contribution to magnetic
field sensitivity is
\begin{equation}\label{Eq:phtnshotnoise}
    \delta B_{psn} = \frac{1}{g_s\mu_BP_z}\frac{
    2(R+\Gamma_{pr}+\Gamma_{SD})}{ l r_e c f n D(\nu) \sqrt{\Phi_0l^2 t}}.
\end{equation}
This can be rearranged
\begin{equation}
    \delta B_{psn} = \frac{1}{g_s\mu_B P_z\sqrt{nVt}}
    \frac{ 2(R+\Gamma_{pr}+\Gamma_{SD}) }{ \sqrt{ \Gamma_{pr}{\rm OD_0}} }\label{Eq:phtnshotnoise2},
\end{equation}
where ${\rm OD_0}=2 r_e c f n l/\Delta\nu$ is the optical depth on
resonance and
\begin{equation}
\Gamma_{pr}=\Phi_0 r_e c f\frac{\Delta\nu/2}{(\nu-\nu_0)^2}
\end{equation}
is the probe rate for far detuned light.

Adding spin-projection noise Eq. \eqref{Eq:atomicshotnoise} and
photon shot noise Eq. \eqref{Eq:phtnshotnoise2} in quadrature,
yields
\begin{eqnarray}
 \delta B &=& \frac{1}{g_s\mu_B
P_z\sqrt{nV t}}\label{Eq:dBtot}\\
\nonumber
&\times&\sqrt{2(R+\Gamma_{pr}+\Gamma_{SD})+\frac{4(R+\Gamma_{pr}+\Gamma_{SD})^2}{\Gamma_{pr}{\rm
OD_0}}}.
\end{eqnarray}
Inspection of Eq. \eqref{Eq:dBtot} shows that the probe pump rate
can be increased until $\Gamma_{pr}\approx\Gamma_{SD}$ without
significant increase to either spin-projection noise or photon shot
noise. If the resonant optical depth ${\rm OD_0}$ is sufficiently
large, the contribution from photon shot noise becomes negligible.
In the limit of infinite resonant optical depth Eq. \eqref{Eq:dBtot}
is optimized for $R=4 \Gamma_{SD}$ and $\Gamma_{pr}=\Gamma_{SD}$ so
that
\begin{equation}\label{Eq:dBopt}
    \delta B_{opt}=\frac{3
    \sqrt{3}}{g_s\mu_B}\sqrt{\frac{\Gamma_{SD}}{nVt}}.
\end{equation}
For a volume of $1~{\rm cm^3}$ and a density of $n=1.7\times
10^{13}~{\rm cm^{-3}}$, and spin-destruction rate
$\Gamma_{SD}=300~{\rm s^{-1}}$ obtained in the experiment at
$103^\circ{\rm C}$, this expression evaluates to about $1.7~{\rm
pG/\sqrt{Hz}}$, where we have assumed that a bandwidth of 1 Hz
corresponds to a measurement time of 0.5 s. In the experiment,
spin-destruction collisions due to alkali-alkali collisions
accounted for only about 1/3 of the total spin-destruction rate, so
somewhat higher sensitivities may be achieved by operating at higher
temperatures where alkali-alkali collisions dominate the
spin-destruction rate.

The above analysis indicates that optimal sensitivity is achieved
when the probe is tuned sufficiently far from resonance so that the
medium is optically thin, minimizing photon shot noise.  Practical
limitations such as finite light power will limit how far one can
detune from resonance. Dropping the assumption of large probe
detuning one finds, for the parameters described above and probe
light tuned 3 optical linewidths away from resonance so that the
optical depth is about 1/3, that the sensitivity is roughly a factor
of 2 larger than the spin-projection noise limit. Of course,
technical sources of noise due to, e.g. vibrations or air currents,
are often much larger than photon shot noise. In this case, it is
desirable to tune the laser closer to optical resonance so that the
optical rotation due to small magnetic fields is larger than other
sources of noise. Finally, we note that the optimal tuning of the
probe light depends on the particulars of the probing scheme.  For
example, Ref. \cite{Rochester2002} considers the case of nonlinear
magneto-optical rotation, where optimal sensitivity is achieved when
the probe is tuned so that there is roughly one optical depth. The
primary reason for the difference is that optical rotation in that
case scales (similarly to absorption) as the inverse square of the
detuning.

\end{document}